# Application of ERA5 and MENA simulations to predict offshore wind energy potential

Shahab Shamshirband [1,2], Amir Mosavi [3], Narjes Nabipour [4], Kwok-wing Chau [5]

[1] Department for Management of Science and Technology Development, Ton Duc Thang University, Ho Chi Minh City, Vietnam

[2] Faculty of Information Technology, Ton Duc Thang University, Ho Chi Minh City, Vietnam

[3] School of the Built Environment, Oxford Brookes University, Oxford OX3 0BP, UK

[4] Institute of Research and Development, Duy Tan University, Da Nang 550000, Vietnam;

[5] Department of Civil and Environmental Engineering, Hong Kong Polytechnic University, Hong Kong, China

**Abstract:** This study explores wind energy resources in different locations through the Gulf of Oman and also their future variability due climate change impacts. In this regard, EC-EARTH near surface wind outputs obtained from CORDEX-MENA simulations are used for historical and future projection of the energy. The ERA5 wind data are employed to assess suitability of the climate model. Moreover, the ERA5 wave data over the study area are applied to compute sea surface roughness as an important variable for converting near surface wind speeds to those of wind speed at turbine hub-height. Considering the power distribution, bathymetry and distance from the coats, some spots as tentative energy hotspots to provide detailed assessment of directional and temporal variability and also to investigate climate change impact studies. RCP8.5 as a common climatic scenario is used to project and extract future variation of the energy in the selected sites. The results of this study demonstrate that the selected locations have a suitable potential for wind power turbine plan and constructions.

1. Introduction

Wind turbine design and installation in offshore and onshore regions are increasingly being used as a suitable proxy for wind energy extraction. As a renewable energy resources, it has been attracted attention of many researchers to develop and investigate its feasibility for different places across the world. However, due to non-uniform spatial and temporal distributions of wind power, regional studies to assess the potential of the wind power for the area under consideration. Therefore, its potential, sustainability, spatio-temporal distributions and economic justification are main factors to develop wind farms for a region. On the other hand, climate change and global warming as global concerns are going to affect many different atmospheric and oceanic parameters in which they have to be taken account for any future plan and development. Wind speed as an important atmospheric variable may experience severe fluctuations and changes under future climatic conditions. Therefore, to achieve a sustainable development and also to enhance reliability of the energy supply from wind power, its spatio-temporal distribution and future variability under climate change impacts should be investigated thoroughly.

Dealing with wind energy exploitation from onshore and offshore wind turbines, wind speed is the most important variable determining magnitude and efficiency of the turbine yields. However, there are some other factors likely to be secondary in importance in which they can change wind energy potential and extraction by influencing the wind turbines. Variation in wind direction, icing, air density, and turbine aging are among the factors affecting efficiency of the turbine outputs beside the magnitude of wind speed. Schindler and Jung (2018) investigated dependence of wind energy to the direction in Germany by means of Copula based approach. It was found that minor changes in wind direction may lead to significant variation in turbine energy yield. Icing is an important factor in high altitudes and arctic latitudes in which in Sweden, Finland, and Norway, the icing or low air temperature reduces turbine operation hours or even though the turbine downtime or stoppage may be attributed to low air temperature (Laakso et al., 2003). Due to inverse relationship between air density and temperature, the higher temperature may decrease wind energy outputs where changing air temperature from 5 to 10 °C can make a decrease about 1 to 2% in air density and to some extend the wind energy (Pryor & Barthelmie, 2010; Solomon et al., 2007). Considering effect of age on wind farm performance, it was found that wind turbines lose about 1.6% of their output per year due to ageing and subsequently, it is expected to decline about 12% of the in wind farm's outputs during a 20 year period of its lifetime (Staffell & Green, 2014).

As the wind speed is the principal component of wind energy derivation, therefore, its sustainability and spatio-temporal variation should be explored appropriately prior to any design and development of wind farms. On the other hand, climate change due to global warming and increasing in greenhouse gas emissions are going to affect many different atmospheric variables in which wind speed is among them. Therefore, investigation of climate change impacts on the energy resources in the study area is a substantial step toward sustainable development and efficient exploitation of wind energy for a longer period. Moreover, effect of climate change on a climatic variable may differ remarkably due to its high spatial variability. In other words, the climate change may lead to increase in wind speed in a region while for another region its impact may be inverse. In this regard, regional analysis of wind climate and its future projection are commonly used. There are many different atmosphere-ocean global circulation models (AOGCMs) projecting climatic variable for historical and future periods based on different scenarios indicating future probable conditions. However, these models are run globally failed to capture local topography and suffer from coarse resolution. Thus, a bias correction or downscaling of the AOGCM outputs is needed to bring simulated capacity factors in line with reality (Staffell & Pfenninger, 2016). The downscaling process is usually carried out by means of statistical and dynamical methods in which the former method has advantages of simplicity and cost effectiveness while the latter one is theoretically preferable to empirical techniques because of its physically consistency with the external conditions (Pryor & Barthelmie, 2010). Among different statistical downscaling techniques, Weibull based methods have taken more popularities that the other techniques and have been successfully applied for this purpose (Alizadeh, Kavianpour, Kamranzad, & Etemad-Shahidi, 2019; Curry, van der Kamp, & Monahan, 2012). However, the coarse resolution of AOGCMs remains unresolved when statistical methods are employed even though they tend to provide a bias corrected outputs. In this regards, regional climate models using dynamical methods are run taking boundary conditions from the global models can represent higher resolution simulations of the variable.

Recently developed regional climate model called coordinated regional climate downscaling experiment (CORDEX) gains advantages of both statistical and dynamical downscaling techniques to provide high resolution climate simulations. Moreover, these simulations are freely accessible which facilitates climate change studies for different regions. They have been widely to project different climatic variables such as solar radiation (Bartók et al., 2017; Frank et al., 2018), wind speed and energy (Ganea, Mereuta, & Rusu, 2018; Moemken, Reyers, Feldmann, & Pinto, 2018) among the others under future scenarios. Considering CORDEX outputs for the European domain to evaluate climate change impacts of the wind power resources, an overall decreasing trend in future power was projected for much of the European domain while for the Black Sea region, there was no remarkable decrease under future climatic conditions (Davy, Gnatiuk, Pettersson, & Bobylev, 2018). For the African domain, a mild increase was projected for future wind climate and energy (Célestin, Emmanuel, Batablinlè, & Marc, 2019; Rautenbach & Herbst, 2016). Also, a decrease in future wave height in the north eastern Atlantic and an increase in significant wave height in the Western part of the Norwegian Sea was projected under future periods. Since the winds are the main drivers of waves, therefore, a similar trend but different in magnitude for the regions can be obtained (Aarnes et al., 2017). However, the literature reveals that wind speed and its future variability have strong sptaio-temporal distribution indicating essence of regional studies for future plan and development. Finally, following the different influencing factors on wind power, it can be derived that wind speed has the highest impact on the wind power among the other factors.

The main purpose of this study is therefore to explore potential of wind energy and its variability under future changing climate for the tentative hotspots. The Oman Sea as a low latitude region was selected as a case study. Due to its warm climate and lack of icing and extremely low weather conditions, it is expected that other factors except the wind speed has the least impacts. Therefore, the study is only focused on wind speed variation and impacts on the wind power for present and future scenarios. In this regard, near surface wind components of CORDEX-MENA outputs for the EC-EARTH model for historical and future periods are obtained and extrapolated for wind power calculations. Moreover, ERA5 wind data for the historical period are applied to evaluate efficiency and accuracy of the RCM. To convert near surface wind speed to those of turbine hub-height, a range of parameters are required such as surface roughness. Due to dependency of roughness length to the wave characteristics in offshore and onshore regions, an special care using the most recent updates on reanalysis waves have been devoted to find the variable. Finally, variability of the wind energy and its spatio-temporal distribution are analyzed for each energy hotspots. The paper is organized in different sections of materials and methods, results and discussion and conclusions in which they are described as follows.

## 2. Materials and methods
### 2.1. Study area

In this research, the Oman Gulf was selected as the case study which is a region of northern Indian Ocean connected to the Persian Gulf through Hormoz Strait. The Oman Gulf is bounded by Iran and Pakistan on north, India on east, and Oman on west. Overall, the study area is located in a low latitude region and roughly with very deep water in which plays an important role in economics of the countries surrounded with. In this regard, many industries, ports, and cities have been developed and they are rapidly growing which increase demand for energy supply. Therefore, it is necessary to plan and construct renewable energy resources to meet sustainable development requirements. Considering potential of the area for renewable energy resource development, wind power can be recognized among the best options to build wind farms in onshore or offshore locations. Due to placing in a low latitude area and having a relatively hot climate, there is not icing problem for turbine blades in which it is a challenge of turbine downtime in northern European wind farms. Moreover, it is a deep water body expecting less disturbance in wind and surface roughness because of topography. However, importance of climate change is not covered on future wind climate and it should be taken under consideration prior any serious practical applications. In this regards, in this study, impacts of climate change on wind energy resources are investigated with a focus on the northern part of the Oman Gulf. However, the offshore areas of the Gulf are mainly very deep in which it is not economic to build wind turbine due to its high construction cost. In this regard, in selection of locations for energy exploitation, both wind power potential and economic aspects (e.g., bathymetry) should be taken under consideration. Figure 1 illustrates location of the study area and its bathymetry. It should be noted that the bathymetry data have been obtained from the website of General Bathymetric Chart of the Oceans (GEBCO).

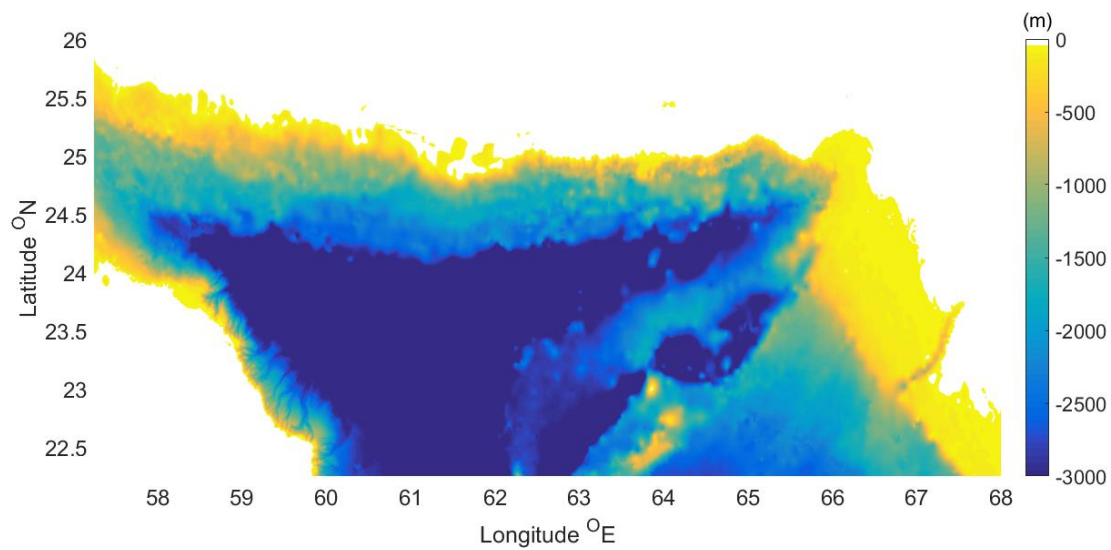

Figure 1. Bathymetry of the study area

## 2.2. Data resources

Datasets play an important role in evaluation of wind energy and climate change impacts. However, a main concern is to convert the near surface wind speed to those of turbine hub-height. Overall, there is rarely measurements of wind speed at the high altitude and most of the meteorological observations are limited in the near surface measurements or ground based masts. In this regard, it is common to validate and consider near surface wind data and subsequently to extrapolate them to the desired height. Thus, in this study, near surface wind speed at 10 m of regional climate model from CORDEX-MENA domain have been obtained to assess wind energy for historical period and for climate change impact studies as well. Overall, CORDEX provides climate data for historical and future scenarios over 14 domains in which the study area is well covered by the Middle East and North Africa (MENA) domain. There are different RCMs for each domain in which they have been derived by using their boundary conditions from different GCMs. In this study, CORDEX-MENA outputs of near surface wind components obtained from EC-EARTH with the spatial resolution of 0.22×0.22 degrees and daily temporal resolution are employed for a 20 year period of 1981-2000 as the historical period. Moreover, the same variable as the historical period for the pessimistic representative concentration pathways (i.e., RCP8.5) for 2081-2100 have been used to project future variation of wind power.

Prior to convert wind speed to turbine hub-height and to compute the wind power, it is necessary to check and evaluate consistency of near surface wind components of the RCM with those of the reference dat. Usually, the field measurements are considered as the reference data, however, for the such a large area, there is only a limited number of stations. In this regard, employing reanalysis data using data assimilation to provide distributed data from numerical models, satellites, and field observations are among the most common procedure in climate change studies. Therefore, they have been widely applied for different purposes by many researchers to evaluate wind climate for different areas (Bednorz, Półrolniczak, Czernecki, & Tomczyk, 2019; Kim, Kim, & Kang, 2018; Rodrigo, Buchlin, van Beeck, Lenaerts, & van den Broeke, 2013; Shanas & Sanil Kumar, 2014). ERA5 as the most recent update on ECMWF (European Center for Medium-Range Weather Forecasts) provides hourly estimates of a large number of atmospheric, land and oceanic climate variables. They gain higher spatial and temporal resolutions compared to their former version of datasets called ERA-Interim. Therefore, near surface wind components at 10 m with spatial resolution of 0.25×0.25 degrees of ERA5 have been used to validate and to assess consistency of the wind data obtained from the RCM. Figure 2 shows mean wind speed at 10 m for historical period of 1981 to 2000 for both ERA5 and EC-EARTH model (the RCM).

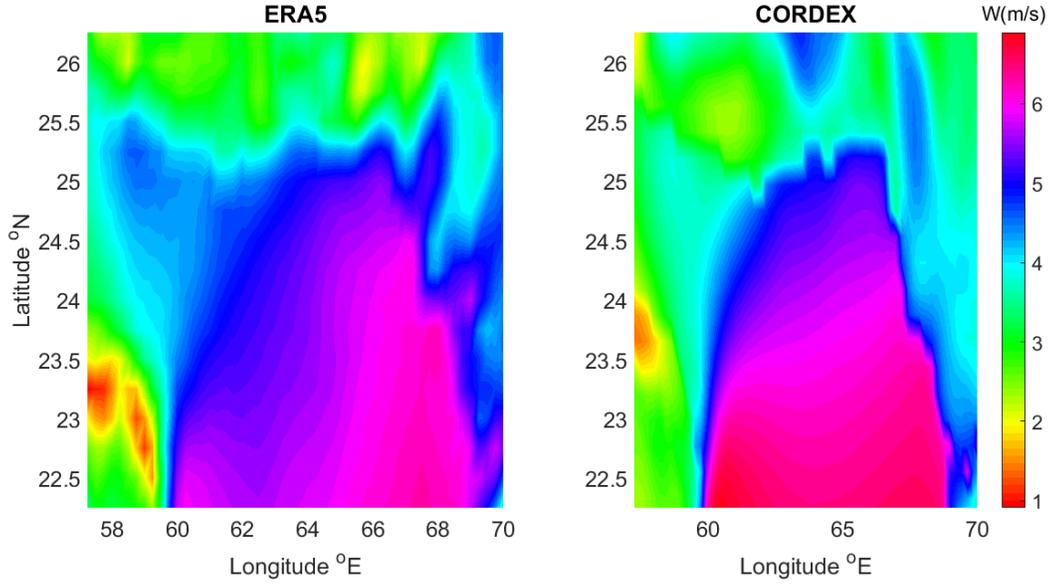

Figure 2. Mean near surface wind speed for a 20 year period of 1981-2000 a) ERA5, b) RCM

As illustrated in Figure 2, there is a good agreement between wind speeds of ERA5 with those of projected by CORDEX-MENA simulations of EC-EARTH. This consistency is associated with both magnitude and spatial distribution of mean wind speed over a 20 year period of 1981-2000. However, it can be observed that ERA5 data slightly overestimate wind speed compared to the RCM. Finally, it can be derived that the CORDEX data can be suitably employed to project and explore future wind climate and wind power due to their appropriate consistency with those of ERA5 here used as the reference data. Subsequent to consistency check of near surface wind speeds, the wind speed have to be converted to turbine hub-height which is usually from 90 to 120 m. This is carried out through a complicated extrapolation process since it is depended on the local conditions such as air density and surface roughness length. To convert surface wind to turbine hub-height, surface roughness length should be determined carefully.

## 2.3. Sea surface roughness

Sea surface roughness is an important parameter affecting atmospheric interactions with sea surface. The momentum transfer between air and sea surface is strongly dependent on sea surface roughness while it is expected that the roughness length to be a function of sea state. It was found that swell and shoaling can remarkably affect roughness where they may cause significant decrease and increase in roughness respectively (Taylor & Yelland, 2001). Therefore, many efforts have been devoted to relate roughness length to sea state measurable variables such significant wave height and wave length. Generally, the roughness length as a function of sea state parameters can be empirically formulated as:

$$\frac{z_0}{H_s} = A\left(\frac{H_s}{L_p}\right)^B \qquad (1)$$

where roughness length is denoted by $z_0$, significant wave height and peak wavelength for combined sea and swell waves are represented by $H_s$ and $L_p$, respectively. $A$ and $B$ are constant coefficients which are determined from experimental or field observations. Following the previous studies, it can be found that 1200 and 4.5 are appropriate estimates for $A$ and $B$, respectively. These values were found to suitably predict magnitude of the drag coefficient for open oceans, lakes, and tanks (Taylor & Yelland, 2001). As wave measurements are mainly reported in terms of significant wave height and wave period, therefore, dispersion equation can be employed to relate wave length ($L$) to wave period ($T$) as:

$$\omega^2 = kgtanh(kh) \qquad (2)$$

where $\omega = 2\pi/T$, $k = 2\pi/L$, g= acceleration gravity and h= water depth. For deep water, $tanh(kh)$ is assumed unit as the water depth is frequently higher than wave length. Thus, for deep waters, dispersion equation can be simplified to calculate wave length as only a function of the wave period as:

$$L = 1.56T^2 \qquad (3)$$

As explained earlier, to compute surface roughness length, significant wave height and wave period data should be acquired for the study area. In this regard, and due to lack of enough filed measurements in the area, ERA5 wave data with resolution of 0.5×0.5 degrees and with one hour interval for a 20 year period of 1981 to 2000 are obtained. It is noticed that the data representing significant wave height are combined sea waves and swells. Subsequent to calculation of surface roughness for each hour, the roughness length was averaged over the whole computational time for the selected stations. Figure 3 depicts mean surface roughness length over the area.

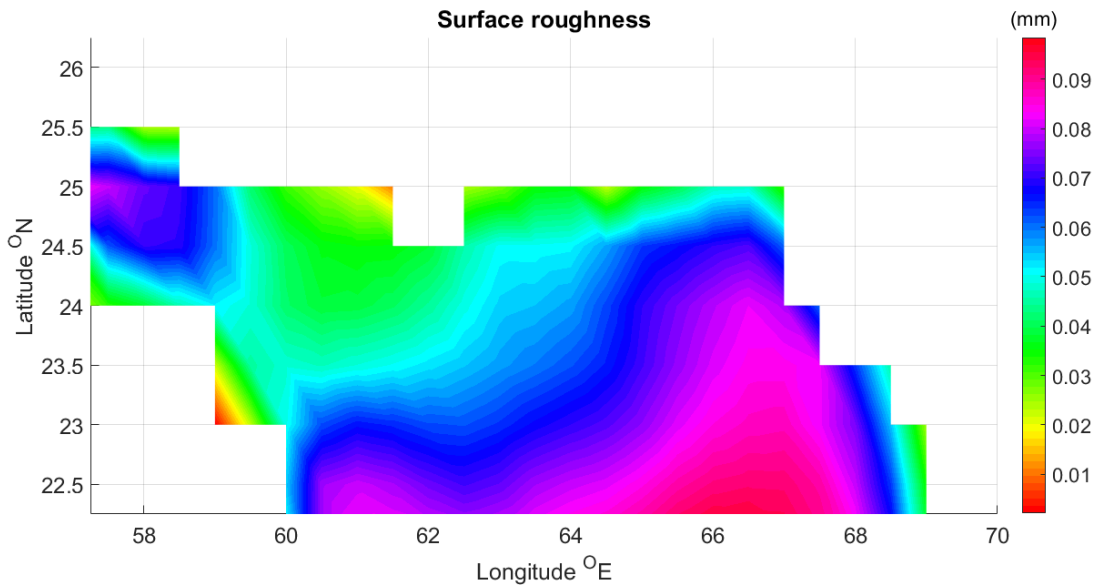

Figure 3. Sea surface roughness length for the study area averaged for a 20 year period

As observed from Figure 3, the mean roughness length for the whole area is less than 0.2 mm while it is common to put a lower limit on roughness length as $z_0 \leq 0.2 \; mm$. In this regard, in our all calculations, the roughness length is considered as 0.2 mm.

**2.4. Weibull distribution and wind energy**

Dealing with wind data, it is common to fit a probability function to evaluate their distribution over a period of time. In this regard, it was found that Weibull distribution function fit suitably for most of the case and it has been widely for wind data analysis. For the study area, it was found that the wind data in Oman Sea follow Weibull distribution accordingly (Malik & Al-Badi, 2009; Sulaiman et al., 2002). Generally, a Weibull distribution function can be mathematically expressed as:

$$f(v) = \frac{k}{c} \left(\frac{v}{c}\right)^{k-1} \exp\left[-\left(\frac{v}{c}\right)^k\right] \tag{4}$$

where c and k called scale and shape parameters of the distribution and they can be determined from the maximum likelihood method as (Chang, Chen, Tu, Yeh, & Wu, 2015):

$$k = \left(\frac{\sum_{i=1}^{n} v_i^k \ln(v_i)}{\sum_{i=1}^{n} v_i^k} - \frac{\sum_{i=1}^{n} \ln(v_i)}{n}\right)^{-1} \tag{5}$$

$$c = \left(\frac{1}{n}\sum_{i=1}^{n} v_i^k\right)^{1/k} \tag{6}$$

where *n* stands for the number of samples.

Alternatively, standard deviation method can be employed to derive Weibull parameters as (Ouammi, Sacile, Zejli, Mimet, & Benchrifa, 2010):

$$k = \left(\frac{\sigma}{\bar{v}}\right)^{-1.086} \tag{7}$$

$$c = \frac{\bar{v}}{\Gamma\left(1+\frac{1}{k}\right)} \tag{8}$$

where $\sigma$ and $\bar{v}$ are standard deviation and average of wind speed, respectively. This study is not to delve in detailed descriptions of calculations of these values and to get through explanations on them, one is referred to Mohammadi, Mostafaeipour, and Sabzpooshani (2014). However, it should be noted that Weibull parameters can be easily computed using MATLAB libraries as for this study. Finally, having the Weibull distribution parameters, the wind power density (P) per unit area (A) can be obtained as:

$$\frac{P}{A} = \frac{1}{2}\rho \int_0^\infty v^3 f(v) dv = \frac{1}{2}\rho c^3 \Gamma\left(1+\frac{3}{k}\right) \tag{9}$$

where $\rho$ denotes for the air density. To reduce computational cost, equation 9 can be replaced with a more simple equation as 10 when wind speed at turbine hub-height is available.

$$P = \frac{1}{2}\rho V^3 \tag{10}$$

where $\rho$ is about 1.225 kg/m³ for the standard conditions and V is the wind speed at the turbine hub-height (z) which can be derived from near surface wind speed at 10 m ($V_{10}$) as:

$$\frac{V}{V_{10}} = \frac{\ln(\frac{z}{z_0})}{\ln(\frac{z_{10}}{z_0})} \tag{11}$$

where $z_0$ is surface roughness length as described in the previous subsection. The usual range for turbine hub-height is from 90 to 120 m in which in this study we considered 120 m for our computations. The last step dealing wind power extraction is to consider turbine cut-in and cut-out wind speed in which is determined from power curve of the turbine. The cut-in and cut-out velocity are the velocities in which no energy output is produces below cut-in and over cut-out velocities. In this study, cut-in and cut-out velocities were considered as 3.5 m/s and 25 m/s, respectively and wind power for velocities below or over these values are excluded in the computations.

### 2.5. Selection of wind energy hotspots

This study was mainly organized to explore climate change impacts on wind power distribution in the northern part of Oman Gulf. As depicted by bathymetry map in Figure 1, the study area is mainly located in deep water and it is expected that the local conditions such as topography has the least disturbance on the wind climate. Moreover, projection of mean wind speed over the area show there is not a rapid gradient. Therefore, to provide more details of climate change impacts, it can be inferred that selecting representative points as the whole. Moreover, the selecting points should be chosen in a way to take account bathymetry, distance to coasts, and magnitude of the wind speed. Furthermore, an attempt is made to consider border line in which for any country bounding the Gulf, at least, there should be a representative point to find climate change impacts over there. In this regard, five locations as the tentative energy hotspots were selected with two points in India, and one for Iran, Pakistan, and Oman. Specifications of the selected points are given in Table 1. The mean wind speed in Table 1 is representative of a 20 year average from 1981 to 2000 derived from ERA5 hourly dataset.

Table 1. Specifications of the selected energy hotspots across the Oman Gulf

| Point No. | Longitude | Latitude | Country | Depth (m) | Mean wind speed (m/s) |
|---|---|---|---|---|---|
| | | | | | |

| P1 | 59.85 | 22.5  | Oman     | 60 | 4.46 |
|----|-------|-------|----------|----|------|
| P2 | 60.75 | 25.25 | Iran     | 45 | 4.06 |
| P3 | 66.75 | 24.75 | Pakistan | 28 | 5.57 |
| P4 | 68    | 23    | India    | 30 | 6.11 |
| P5 | 68    | 22.5  | India    | 70 | 6.19 |

Finally, the climate change impacts on wind power for the selected points are taken under consideration in terms of mean annual wind power, seasonal variability, and directional distribution of the power for the historical and future scenario of RCP8.5. Dealing with climate change impacts on wind power, it is noticed that mean values are more interesting. On the other hand, extreme values (the speeds higher than cut-out or lower than cut-in wind speeds) have the least importance since they are excluded in the computations.

## 3. Results and discussion

The mean annual distribution of wind power for historical and future periods under climatic scenario of RCP8.5 have been projected over the whole domain. Moreover, the relative change in wind power in future compared to 100 years ago have been calculated to highlight climate change impacts in the study area. Afterwards, detailed investigations on the selected sites as tentative energy hotspots are presented. Figure 4 projects mean annual wind power for historical (1981-2000) and future (2081-2100) periods to illustrate climate change impacts on the energy resource in the study area.

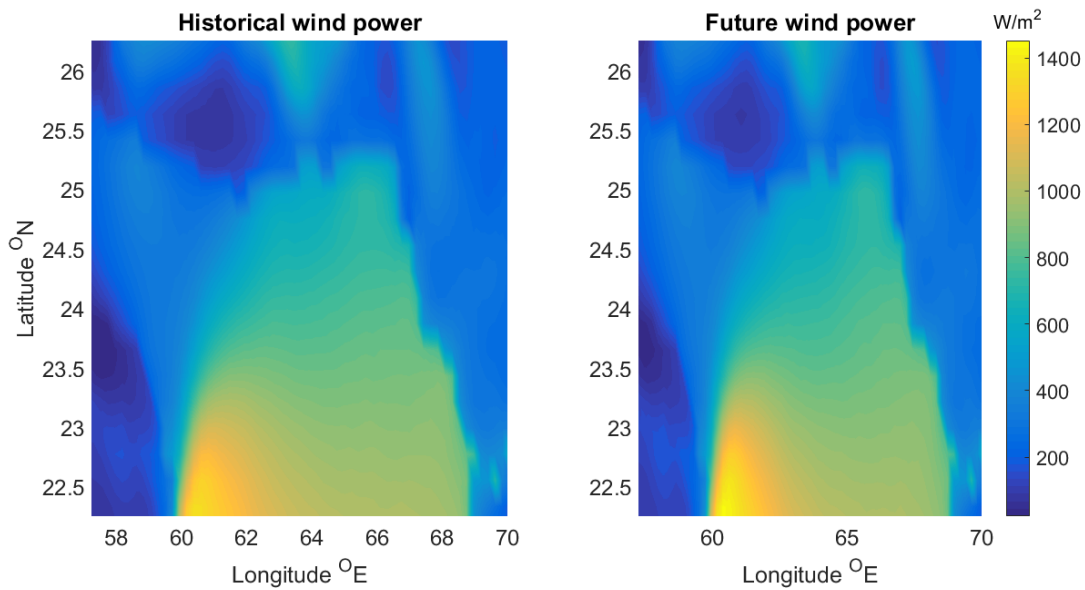

Figure 4. Mean annual variation of wind power a) historical period, b) Future period

As observed from Figure 4, there is strongly spatial variability in wind power distribution over the Oman Gulf. In the northern stripe of the Gulf, a relatively low potential of wind power (less than 200 or 300 W/m$^2$) representing that a roughly calm state is governed on the region. On the other hand, it can be found that the wind power exceeds 1000 W/m$^2$ indicating high potential for wind power. Generally, it can be derived that moving from northern parts of the Gulf towards southern regions, wind power tends to increase significantly. Moreover, it can be seen that the power is stronger in the western of the Gulf compared to the eastern regions which is mainly due to prevailing wind trend in low latitudes blowing westward. However, high gradients in longitudinal and latitudinal variability in wind power in the study area demonstrate strong spatial variability of the wind power where the mean annual wind power changes in a wide range of 100 to 1400 W/m$^2$. Generally speaking, it can be derived that wind power in Oman has the highest values and In Iran has the lowest values. India and Pakistan can be categorized as the region with relatively intermediate potential for wind power in which the power for India is projected higher than the energy potential in Pakistan. Comparing wind power projections for historical and future periods, it can be derived that the wind power will not change remarkably under future climatic conditions. The historical and future distribution and magnitude of the wind power in the domain demonstrate a similar trend indicating sustainability of this source of renewable energy. However, to provide more details on the climate change impacts on wind climate (wind power variation over 100 years), relative changes representing difference in mean annual wind power in future compared to the historical power are illustrated in Figure 5.

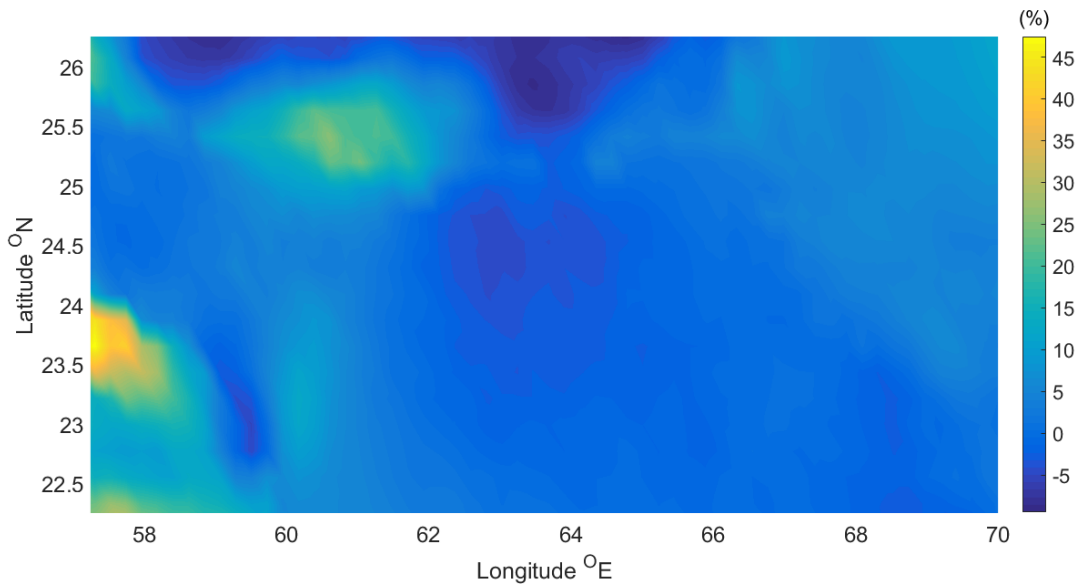

Figure 5. Mean relative changes in wind power

According to Figure 5, it can be concluded that the relative change in wind power for future for the major part of the domain is less than 5 percent regardless of the negative or positive sign. Low values of relative changes projected for the future wind power in the area are promising toward sustainable development of wind farms for energy purposes. The highest increase and decrease in future wind power are expected to happen in the western and northern parts of the Gulf, respectively. Generally, for the offshore and sea areas, it can be derived that climate change has a negative impacts tending to decrease the future wind power potential even though this decrease is not remarkable and rarely exceeds 5% in magnitude. On the other hand, in land areas, especially for the land near to the Gulf in the western of the Gulf, a rapid growth in wind climate is projected for future where the future wind power are about 40% higher than the power in the historical period. However, for the land areas surrounding the Gulf in its eastern part, the trend is still decreasing but its magnitude is frequently lower than the variation rate in the western part of the Gulf. Thus, in the land area adjacent to the Gulf in its eastern part, a mildly increasing trend (ranging from 0 to 10%) is projected for the future period. However, offshore and nearshore areas are mainly expected to experience a decreasing trend under the future climatic conditions although its magnitude is not remarkable. Therefore, considering the relative variations of wind power and its spatial distribution, it can be obtained that climate change is not a menace for sustainability of wind power production in offshore and nearshore areas of the Oman Gulf. Table 2 gives annual and relative changes in the power for the selected points across the Gulf to provide a quantitative analysis of the energy.

Table 2. Mean annual and mean relative change in wind power for RCP8.5

|  | P1 | P2 | P3 | P4 | P5 |
|---|---|---|---|---|---|
| Historical (1981-2000) | 573.37 | 108.64 | 342.67 | 911.23 | 942.56 |
| RCP8.5 (2081-2100) | 611.37 | 132.27 | 357.61 | 898.12 | 927.75 |
| Relative change (%) | 6.63 | 21.75 | 4.36 | -1.44 | -1.57 |

Regarding Table 2, it is obtained that the points 4 and 5 (P4 and P5) located in India have the highest energy potential for both historical and future projections with power higher than 900 Watt per square meter. Moreover, the point located in Oman also showed a suitable potential of wind energy with mean annual power about 600 Watt per square meter. However, the point representing wind power in Iran demonstrates a relatively lower potential among the others. The other point near to point 2 which is in Pakistan (P3) has an intermediate power when compared with 4 other points. Comparing the power for historical and future periods does not stand for a rapid change under future climatic conditions indicating sustainability in wind power potential for the selected locations. This sustainability is approved by very low negative values of relative change in which the projected decrease for points 4 and 5 are not greater than 2%. On the other hand, for the other points which increasing trends are obtained, this increase changes in a range of 4 to 22%. Therefore, it is expected that climate change impacts do not decrease future wind power at least and also for some regions it can enhance the power production. The highest relative change in future wind power is related to the point with the lowest potential (P2) which is desirable for energy extraction since the power under future condition has a relatively high increase in the area. Generally, the results presented in Table 2 are promising toward sustainability in wind energy extraction for the points under consideration as the future climatic conditions are not going to make a remarkable negative change on the wind power distribution. Subsequently, directional distribution of wind power is a key step as a minor change in direction can remarkably affect the wind turbine energy yield. Moreover, the main directions of the winds play an important role in design and arrangement of the wind turbines. In this regard, directional analysis for wind power distribution for the selected points and for historical and future periods are presented as Figure 6 and Figure 7, respectively.

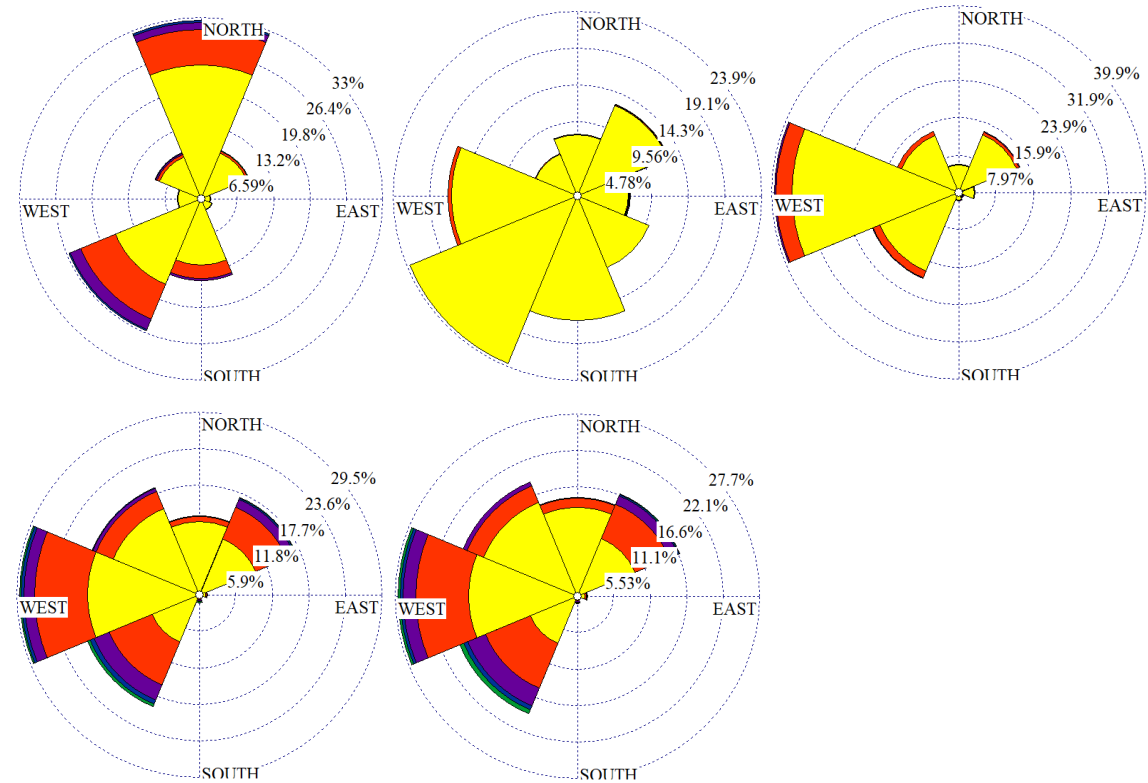

Figure 6. Directional distribution of wind power for the selected stations during historical period

Overall, Figure 6 reveals strong variability in directional distribution of wind power where for the points in the eastern part of Gulf, the dominant winds are blowing from west to east. On the other hand, for the point number 1 where located in Oman (in the western of the Gulf), the dominant winds mainly come from north and southwest. Furthermore, south winds contributes about 10% of the winds in this area. For higher latitudes (points 2 and 3), effects of northern winds are diminished while western and southwestern winds are augmented. For point 2, southern and northeastern winds are ranked as main directions just after western and southwestern winds. For the points in the lower latitudes (points 4 and 5), wind power has a roughly similar directional distribution in which west, southwest, northwest, northeast, and north winds are main contributors of wind directions, respectively. Considering the directional distribution of wind power, it can be derived that for all the points, there is a wide range of variation which necessitates optimum design in turbine arrangement of wind farms to achieve the best efficiency. Generally speaking, such a wide range of variation in wind direction is not desirable for wind power extraction even though the wind farms can be planned in a way to reach the highest rate of production. Dealing with climate change impacts, directional analysis of wind power and also its variability related to those of the historical period are key elements toward achieving sustainable development and optimum efficiency of the wind farms. In this regard, future distribution of wind power for the selected points are illustrated as Figure 7.

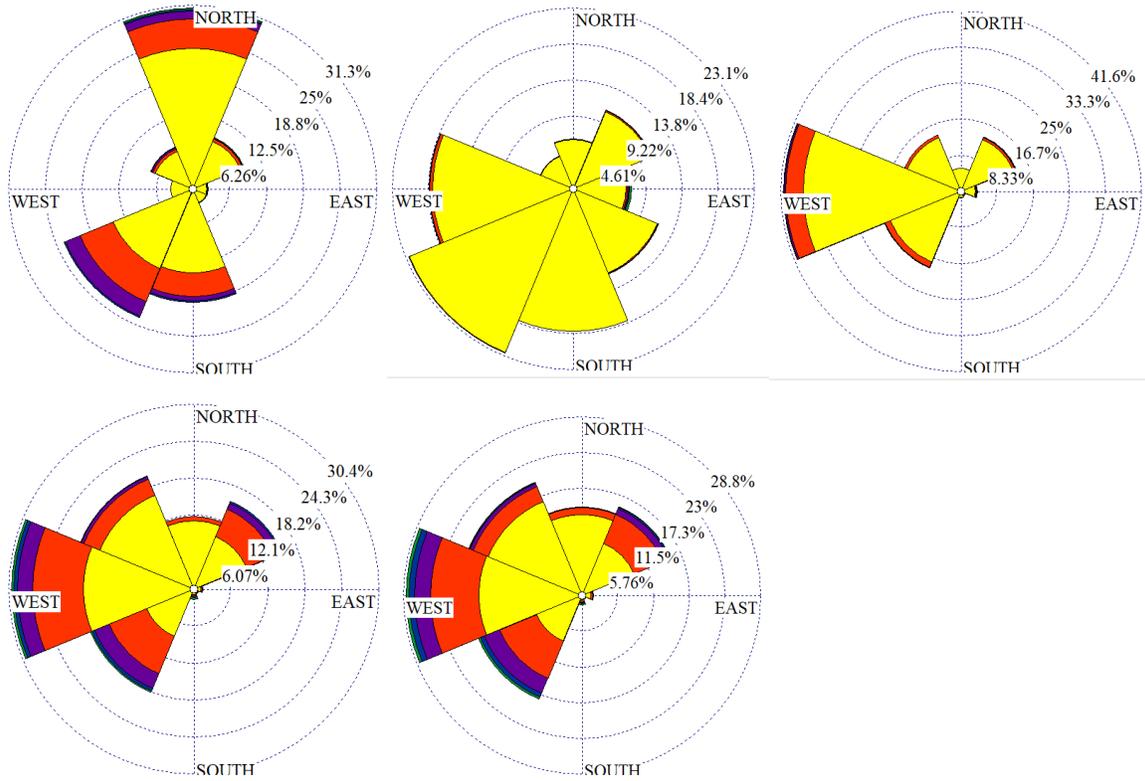

Figure 7. Directional distribution of wind power for the selected stations during future scenario

Regarding Figure 7, similar conclusions as Figure 6 can be derived in which the wind direction has a wide range of variation for different points. Concerning the climate change impacts, it can be found that the main directions for wind power under the future climatic scenario will remain the same as the historical directional analysis but with minor changes in some points in terms of direction and magnitude. Generally, it can be obtained that the climate change does not affect the directional distribution of wind power remarkably. For point 1, the southern winds are augmented in frequency and magnitude under the future warming conditions. For the other points, no significant change in wind directions can be derived. Therefore, directional analysis of wind power under the future climatic scenario is consistent with the sustainability of the wind power for energy extraction and exploitation. However, investigation of temporal variability of the wind power is another key steps to meet demand-supply criterion. In this regard, seasonal analysis of wind power for historical and also its variability under the future scenario is carried out to provide more details on temporal distribution of wind power for the selected points. Table

3 presents result of mean seasonal wind power for historical and future periods in the selected sites.

Table 3. Mean seasonal wind power for historical and future periods

|  |  | P1 | P2 | P3 | P4 | P5 |
|---|---|---|---|---|---|---|
| Winter | His. | 677.05 | 86.14 | 180.21 | 729.21 | 784.31 |
|  | RCP 8.5 | 587.16 | 78.90 | 192.74 | 732.11 | 775.23 |
| Spring | His. | 270.45 | 118.81 | 316.54 | 734.59 | 720.11 |
|  | RCP 8.5 | 301.76 | 129.22 | 335.54 | 759.01 | 738.66 |
| Summer | His. | 906.26 | 147.41 | 598.40 | 1452.28 | 1533.15 |
|  | RCP 8.5 | 1038.92 | 146.16 | 634.75 | 1448.77 | 1526.89 |
| Autumn | His. | 440.24 | 81.47 | 268.87 | 720.39 | 727.31 |
|  | RCP 8.5 | 520.15 | 178.80 | 259.46 | 649.79 | 664.62 |

Concerning the seasonal variation of wind power in the selected points, it can be inferred that summer and autumn are the strongest and weakest seasons, respectively. A high temporal variability in terms of seasonal projection is projected in which the difference in wind power for the strongest and the weakest seasons are significant. In line with the results of annual analysis, the seasonal distribution wind power indicates that points 4 and 5 located in lower latitudes than the others have the highest potential for wind power production. Afterwards, point 1 located in Oman and point 3 located in Pakistan are ranked subsequently. Comparing the results of mean seasonal wind power for historical with those of future projections, the trend is not monotonic and there is a high spatial variability in which in some point the trend is decreasing while increasing for the others. However, the power in spring is expected to increase for all the selected points. The variation of other seasons are strongly dependent on the location under consideration. Therefore, in Figures 8 to 12, the seasonal results for each point are illustrated separately.

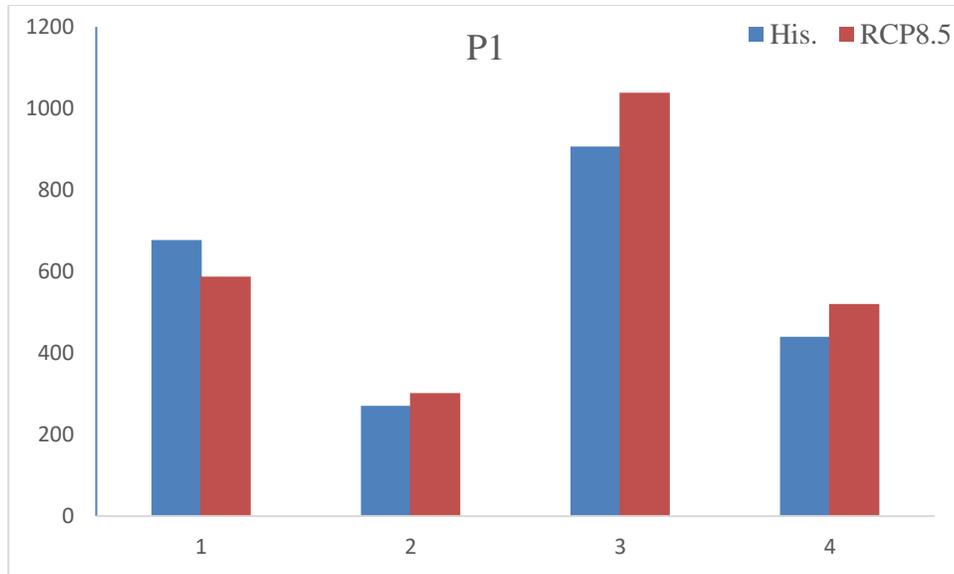

Figure 8. Seasonal variation of wind power for station 1

Figure 8 reveals that wind power for point 1 has increasing trends for all seasons except winter in which the wind power slightly decreases. Moreover, it can be found that summer and spring are among the seasons with the highest and lowest wind power respectively. Overall, it can be inferred that climate change in the selected point do not change wind power remarkably since the future increase or decrease compared to historical values are marginally. The high power potential in summer is promising as the demand for energy in this seasonal is frequently higher than the other seasons. Thus, temporal distribution of wind power are to some extent in line with the demand to energy for the area.

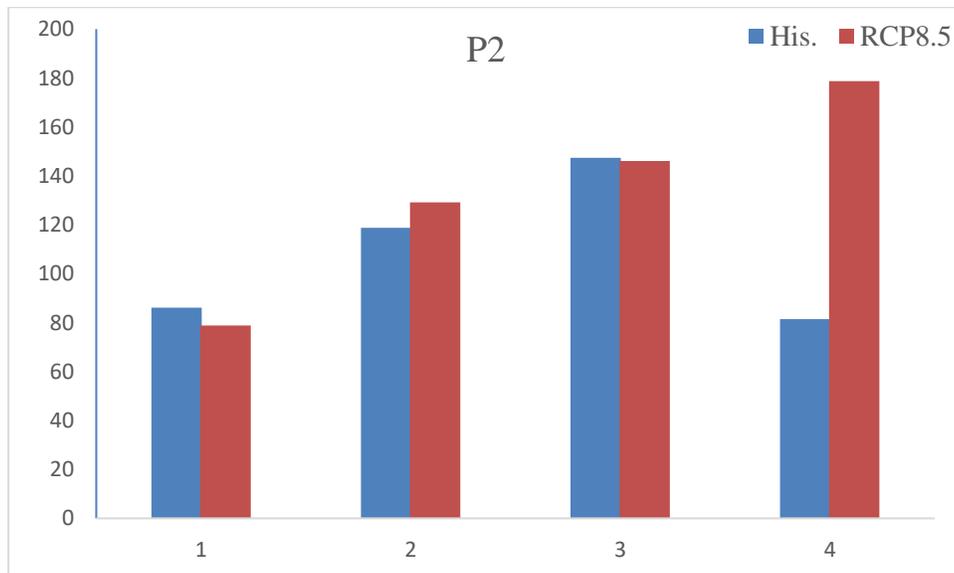

Figure 9. Seasonal variation of wind power for station 2

For point 2 located in Iran, the wind power has the least values among the other points while it does not reach to 200 W/m$^2$ even though for the season with the highest power (autumn). Generally, wind power for future have a roughly similar values of the historical ones except for the autumn in which a rapid increase is obtained. Finally, it should be noticed that the wind power in this point is relatively lower than the other points but still worthy due to essence of renewable energy for future development. Compared to point 1, the season representing the highest and the lowest wind power are completely different in point 2. Therefore, the temporal variation of wind power in terms of seasonal distribution shows a strong dependency on the location and its spatial characteristics.

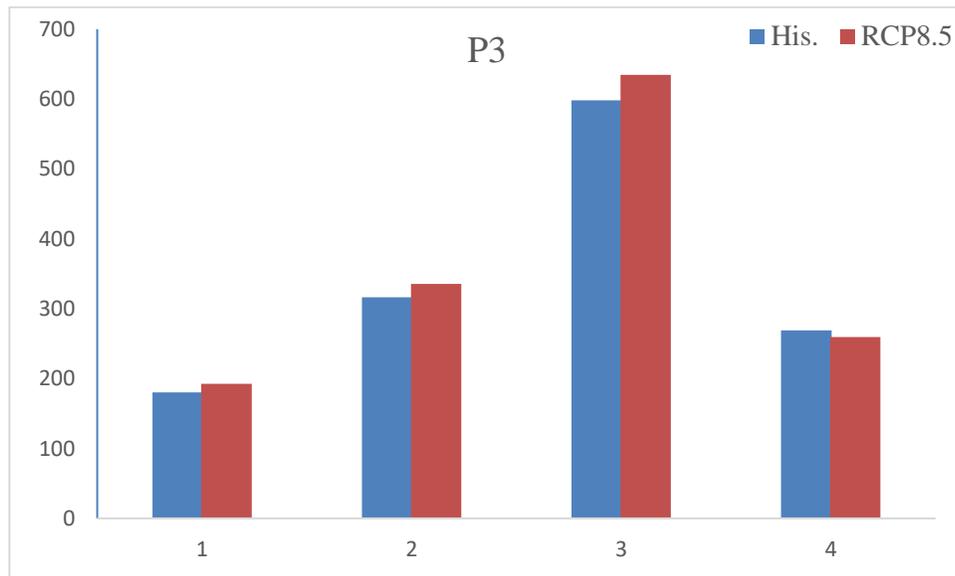

Figure 10. Seasonal variation of wind power for station 3

As observed from Figure 10 representing the results for point 3 in Pakistan, the future projections of wind power do not significantly differ from the corresponding values in the historical period. Moreover, the future values of wind power are slightly higher than the historical values indicating positive impacts of climate change on the wind energy resources. However, for autumn, a decreasing trend in future is observed but its magnitude is negligible. The strongest and weakest seasons are summer and wind which are desirable for energy demand and extraction for the location and future development.

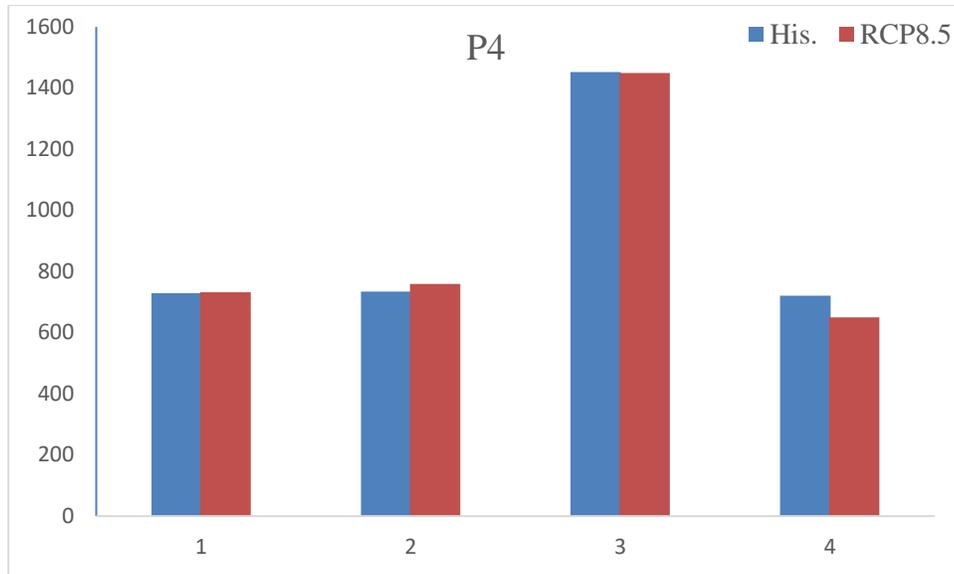

Figure 11. Seasonal variation of wind power for station 4

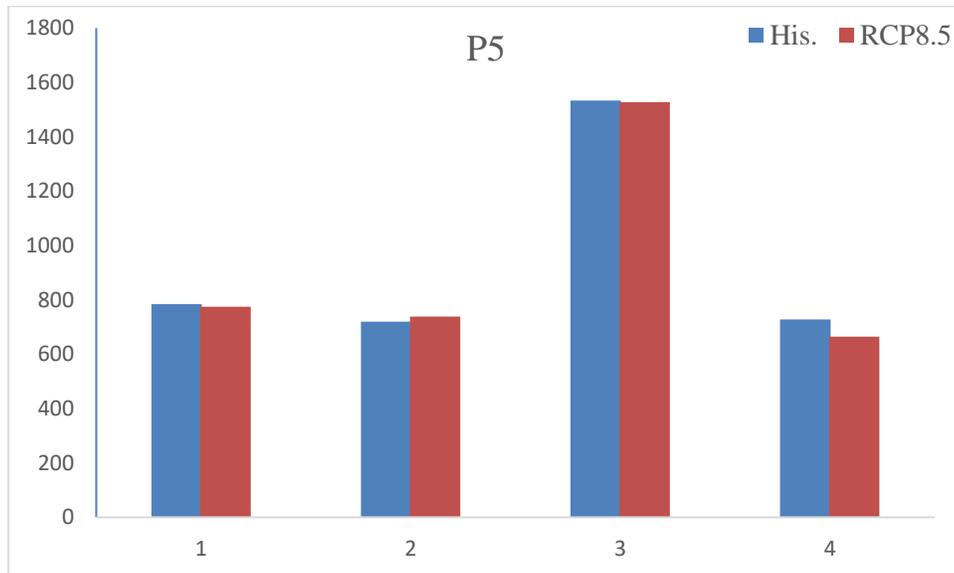

Figure 12. Seasonal variation of wind power for station 5

Figures 11 and 12 illustrate seasonal power distribution for two points located in India along Oman Gulf. These two points which have a similar longitude but a small difference in latitude provide almost a similar seasonal distribution and pattern. For both points, the summer is the strongest season while the autumn is the season with the lowest wind power. Future fluctuations in wind power compared with those of historical period are negligible indicating that climate change do not inversely affect wind energy resources in the area. Moreover, higher wind power potential in summer is of interest due to higher demand for energy in summer.

## 4. Conclusions

Wind power has been recognized among the common types of renewable energy due to its availability and abundance. However, its highly spatial and temporal distributions require further analysis toward achieving the highest efficiency in any future plan and construction as well as in existing wind farms. Recently, climate change impacts on different atmospheric and oceanic phenomena attracted attention of many researchers and many different simulations under different scenarios have been developed. Therefore, having future simulation for wind speed, considering climate change impacts for wind power assessment is necessary to reach a sustainable development and increase lifetime of wind farms. In this study, wind power potential and climate change impacts in Oman Gulf bounded by four countries with high rate of growth in population and industry and consequently with an increasingly demand for power supply have taken under consideration. To do that, a high resolution regional climate model (EC-EARTH) obtained from CORDEX program for the MENA domain to evaluate historical and future variability of the wind power. For future simulations, near surface wind outputs for RCP8.5 have been employed. Prior to convert near surface wind to wind power at turbine hub-height, consistency of the wind outputs obtained from the regional climate model have been evaluated against ERA5 wind data. Moreover, wave outputs obtained from ERA5 have been used to compute sea surface roughness required to data extrapolation to turbine hub-height. The results of wind power in terms of mean annual, mean seasonal and directional distributions have been analyzed for historical and future periods.

Generally, wind power computations for the selected points revealed high spatial and temporal variability but low impacts of climate change on wind power in future period. It was found that the wind power are relatively high for the points located in India and Oman while the point located in Iran has relatively low potential for wind power production. Moreover, it was found that the lower latitudes have higher potential than areas in higher latitudes. Concerning the climate change impacts, no remarkable negative impacts due to climate change was found for the future climatic scenario under consideration. On the other hand, for some points, an increasing trend was obtained. Similarly, the directional analysis of wind power demonstrated that future wind power for the selected points has roughly the same distribution as the historical projections. However, it is noticed that the wind directions have a wide range of variation in which necessitates to design and construct wind farms with optimum arrangements of wind turbines to produce power for wind in different directions not only in the main direction.

Seasonal analysis of wind power projects a high temporal variability for all the points in which summer was found to have the highest energy potential while the weakest season is different point by point. Considering wind power estimations for historical and future scenario, it was derived that the power exceeds 1000 W/m$^2$ for the points in India and Oman indicating a very high potential for the energy extraction. Moreover, future simulations of wind power revealed that the climate change does not affect the power potential inversely which is in line with the results of annual analysis. Finally, it can be concluded that wind power has a suitable potential for the areas considered in this study as its seasonal distribution, future projections and magnitude are all demonstrating sustainability and efficiency of the points for further developments of the energy facilities. In other words, no or minor change in future wind potential, high values of wind power and the highest wind power for summer are all promising as the summer is usually the season with the highest demand for the energy. Moreover, the countries surrounding Oman Gulf are subject to rapid growth in population and industry requiring increasingly demand for future activities in energy.